# Multi-vortex versus interstitial vortices scenario in superconducting antidot arrays

A. D. Thakur[1,*], S. Ooi[1], S. P. Chockalingam[1], J. Jesudasan[2], P. Raychaudhuri[2], K. Hirata[1]

[1]Superconducting Materials Center, National Institute for Materials Science, 1-2-1 Sengen, Tsukuba, Ibaraki 305-0051, Japan

[2]Department of Condensed Matter Physics and Materials Science, Tata Institute of Fundamental Research, Homi Bhabha Road, Colaba, Mumbai 400005, India

Abstract

In superconducting thin films, engineered lattice of antidots (holes) act as an array of columnar pinning sites for the vortices and thus lead to vortex matching phenomena at commensurate fields guided by the lattice spacing. The strength and nature of vortex pinning is determined by the geometrical characteristics of the antidot lattice (such as the lattice spacing $a_0$, antidot diameter $d$, lattice symmetry, orientation, etc) along with the characteristic length scales of the superconducting thin films, viz., the coherence length ($\xi$) and the penetration depth ($\lambda$). There are at least two competing scenarios: (i) multiple vortices sit on each of the antidots at a higher matching period, and, (ii) there is nucleation of vortices at the interstitial sites at higher matching periods. Furthermore it is also possible for the nucleated interstitial vortices to reorder under suitable conditions. We present our experimental results on NbN antidot arrays in the light of the above scenarios.

PACS codes: 74.10.+v, 73.60.Ka, 74.40.+k, 74.80.-g, 74.78.-w, 74.81.Fa, 75.75.+a, 73.21.-b

Keywords: Superconducting film; antidot array; columnar pinning sites; matching effect; NbN

*Corresponding author.

Dr. A. D. Thakur

Postal address: Superconducting Materials Center, National Institute for Materials Science, 1-2-1 Sengen, Tsukuba, Ibaraki 305-0051, Japan

Phone: +81-29-851-3354 (x 6627)

Fax: +81-29-859-2301

E-mail address: adthakur@gmail.com, thakur.ajay@nims.go.jp

## 1. Introduction

Engineered lattice of antidots (holes) in a superconducting film acts as an array of columnar pinning sites for the vortices and thus leads to vortex matching phenomena at commensurate fields guided by the lattice spacing [1]. Extensive works in recent past have established that the vortices form highly ordered configurations at integer and fractional matching fields $H_n = n\Phi_0/S$ and $H_{k/l} = (k/l)\Phi_0/S$ respectively, where, $n, k, l$ are integers, $\Phi_0 = 2.07 \times 10^{-7}$ G-cm$^2$ is the flux quantum and $S$ is the primitive unit cell area of the engineered antidot lattice [2, 3]. The exact nature and strength of vortex pinning is determined by the geometrical characteristics of the antidot lattice (such as the lattice spacing (viz., $a_0$), antidot diameter $d$, lattice symmetry, orientation, etc) along with the characteristic length scales of the superconducting thin films, viz., the coherence length ($\xi$), and, penetration depth ($\lambda$). There are at least two competing scenarios: (i) multiple vortices sit on each of the antidots at a higher matching period, and, (ii) there is nucleation of vortices at the interstitial sites at higher matching periods. Horng et al [4] demonstrated the absence of certain matching periods in the case when interstitial vortices nucleate in the antidot array samples. Furthermore, there exists experimental evidence for a possible reordering of the nucleated interstitial vortices leading to a "super matching (SL)" flux line lattices [5-7]. Within the theoretical scenario proposed by Mkrtchyan and Shmidt [8], the maximum number of vortices captured by an isolated antidot (columnar defect) of diameter $D$ is given by the saturation number ($n_s$) where, $n_s = D/4\xi(t)$, such that $\xi(t) = \xi_0/\sqrt{(1-t)}$, with $\xi_0$ being the zero temperature coherence length and $t$ ($=T/T_c$) being the reduced temperature. However, taking into consideration the vortex-vortex interaction in the case of a lattice of antidots, Doria et al proposed $n_s \sim (D/2\xi(t))^2$ [9]. Recently we demonstrated the suitability of NbN thin films for studying the physics of vortex matching phenomena in engineered antidot lattices [10]. Here, we present our experimental results on a triangular NbN antidot array in the light of the above scenarios.

## 2. Experimental

The antidot array samples were prepared on a Si substrate. For this initially a 60 nm thick NbN film was deposited by sputtering. A probe pattern is then made via photolithography followed by ion beam etching for a typical four probe measurement with region between the voltage probes being 40 μm x 40 μm wide. In these 40μm x 40μm wide available regions, antidot arrays/wire networks in suitable geometries are patterned using the Focussed Ion Beam (FIB) milling technique as reported elsewhere [10-12]. The flux-flow resistance measurements were carried out via the conventional four-probe technique using a home-made insert which goes into the Quantum Design SQUID MPMS XL. The data acquisition was done utilizing the external device control (EDC) option to operate the Keithley 6430 sub-femtoamp remote sourcemeter and the Keithley 2182A nanovoltmeter. The *DLL* files for this purpose were generated using a Delphi program. This way we use the Quantum Design SQUID MPMS XL primarily as a cryostat providing precise temperature and magnetic field control. An additional temperature sensor was placed very close to the sample location in the home-made insert and read using a Cryocon 62 ac resistance bridge also controlled via EDC.

## 3. Results and Discussions

The details of the NbN film used for fabricating the triangular antidot array is shown in Table 1 [13]. The details of the sample fabricated for the purpose of the experiments are shown in Fig. 1. The main panel shows a typical four probe configuration with 10 available regions for fabricating antidot arrays. The inset (shown by an arrow) corresponds to a portion of triangular antidot array fabricated using FIB milling. The lattice parameter is 400 nm and the antidot diameter is 170 nm. Figure 2 shows a plot of flux flow resistance, $R$ versus the filling fraction, $f$ $(=H/H_1$, where $H_1$ is the first matching period) obtained at various drive currents in the range of 10-100μA at a reduced temperature value of 0.937. Robust integer vortex matching effect is observed up to the eighth matching period at a drive current of 10 μA however, at a drive current of 100 μA signatures for matching periods beyond the sixth period were absent. The integer-matching period (= 149.4 Oe) is marked as $H_1$. Measurements were done at various temperatures close to $T_c$ in order to study the effect of temperature on the observed phenomena. The top panel in Fig. 3 shows $R$ versus $f$ data at various reduced temperatures ($t$) in the range 0.931-0.943 close to $T_c$. In the bottom panel is shown the experimentally observed data (filled blue circles) of the saturation number, $n_s$ as a function of the reduced temperature, $t$. Also plotted in the bottom panel are the theoretical estimates for $n_s$ within the scenarios of Mkrtchyan and Shmidt [8] (solid green line) and Doria et al [9] (solid red line). As can be observed, whereas the Mkrtchyan and Shmidt result underestimates the value of $n_s$, the result of Doria et al in its primitive form makes an overestimation. Here one should keep in mind that the simulated annealing method of Doria et al [9] just predicts the variation of $n_s$ with $D/\xi$ (i.e., $n_s \sim (D/2\xi)^2$) and the true determination of the proportionality constant depends on a number of factors, viz., the symmetry of the antidot array lattice, material characteristics, etc which needs to be estimated within a suitable microscopic theory. Recently, Berdiyorov et al [14-16], obtained the equilibrium structural phase diagram showing the different ground-state vortex configurations as a function of the size and periodicity of the antidots in a square array. Another plausible scenario taking into account the interstitial vortex picture, corresponds to a situation where, there is a reordering of the nucleated interstitial vortices leading to a "super matching (SL)" flux line lattices [5-7]. Simulations on triangular antidot array by Reichhardt et al [17] corroborate this scenario.

## 4. Conclusions

We demonstrated robust signatures of vortex matching phenomenon in flux-flow measurements on thin films of NbN with engineered triangular array of antidots. The observed results are discussed within the light of theories by Mkrtchyan and Shmidt [8] and Doria et al [9], and simulations by Reichhardt et al [15].

**Acknowledgements**

ADT and SO would like to acknowledge partial support from World Premier International Research Center (WPI) Initiative on Materials Nanoarchitectonics, MEXT, Japan.

**Table 1:** List of physical quantities related to the NbN film used for fabricating the triangular antidot array sample [13].

| Quantity | Value |
|---|---|
| Transition temperature, $T_c$ | 15.0 K |
| $k_F l$ | 5.9 |
| Resistivity in normal state, $\rho_N$ | 1.48 μΩ-cm |
| Electron mean free path, $l$ | 3 Å |
| Carrier density, $n$ | $1.61 \times 10^{29}$ m$^{-3}$ |
| Coherence length, $\xi(0)$ | 4.3 nm |

**Figure Captions**

**Fig. 1:** The main panel shows a typical four probe configuration with ten available regions for fabricating antidot arrays. The inset (shown by an arrow) corresponds to a portion of triangular antidot array fabricated using FIB milling. The lattice parameter is 400 nm and the antidot diameter is 170 nm.

**Fig. 2:** Plot of R versus f obtained at various drive currents in the range of 10-100μA at a reduced temperature value of 0.937.

**Fig. 3:** The top panel shows R versus f data at various reduced temperatures, t (= $T/T_c$) in the range 0.931-0.943 close to $T_c$. The integer-matching period (= 149.4 Oe) is marked as $H_1$. In the bottom panel is shown the experimentally observed data (filled blue circles) of the saturation number, $n_s$ as a function of the reduced temperature, t. Also plotted in the bottom panel are the theoretical estimates for $n_s$ within the scenarios of Mkrtchyan and Shmidt [8] (solid green line) and Doria et al [9] (solid red line).

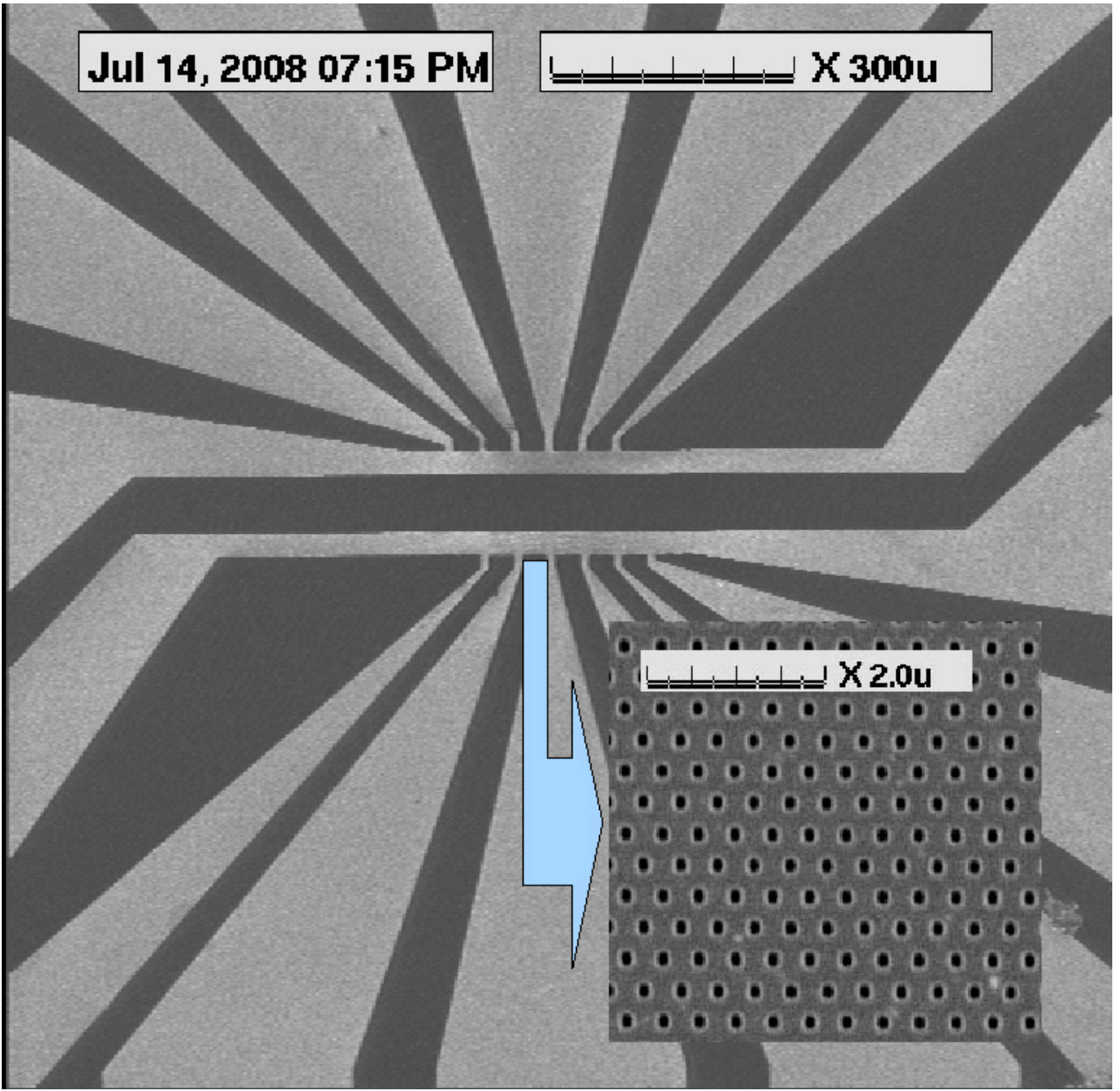

**Fig.1**

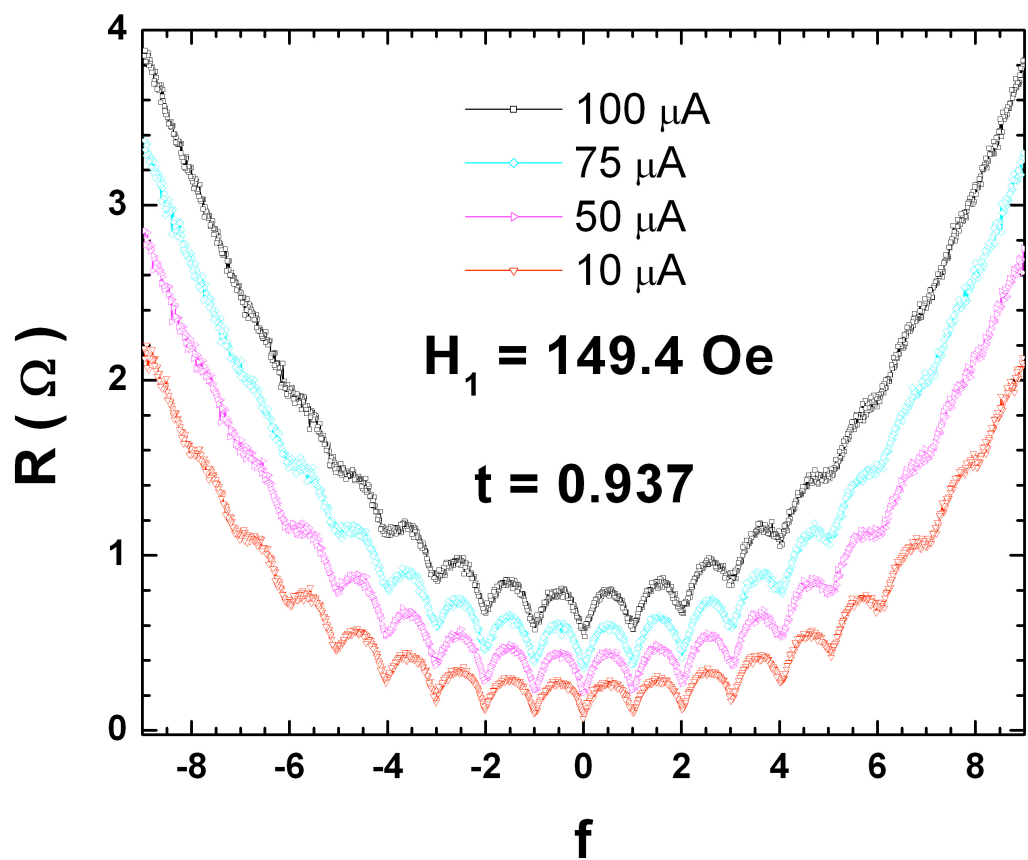

Fig. 2

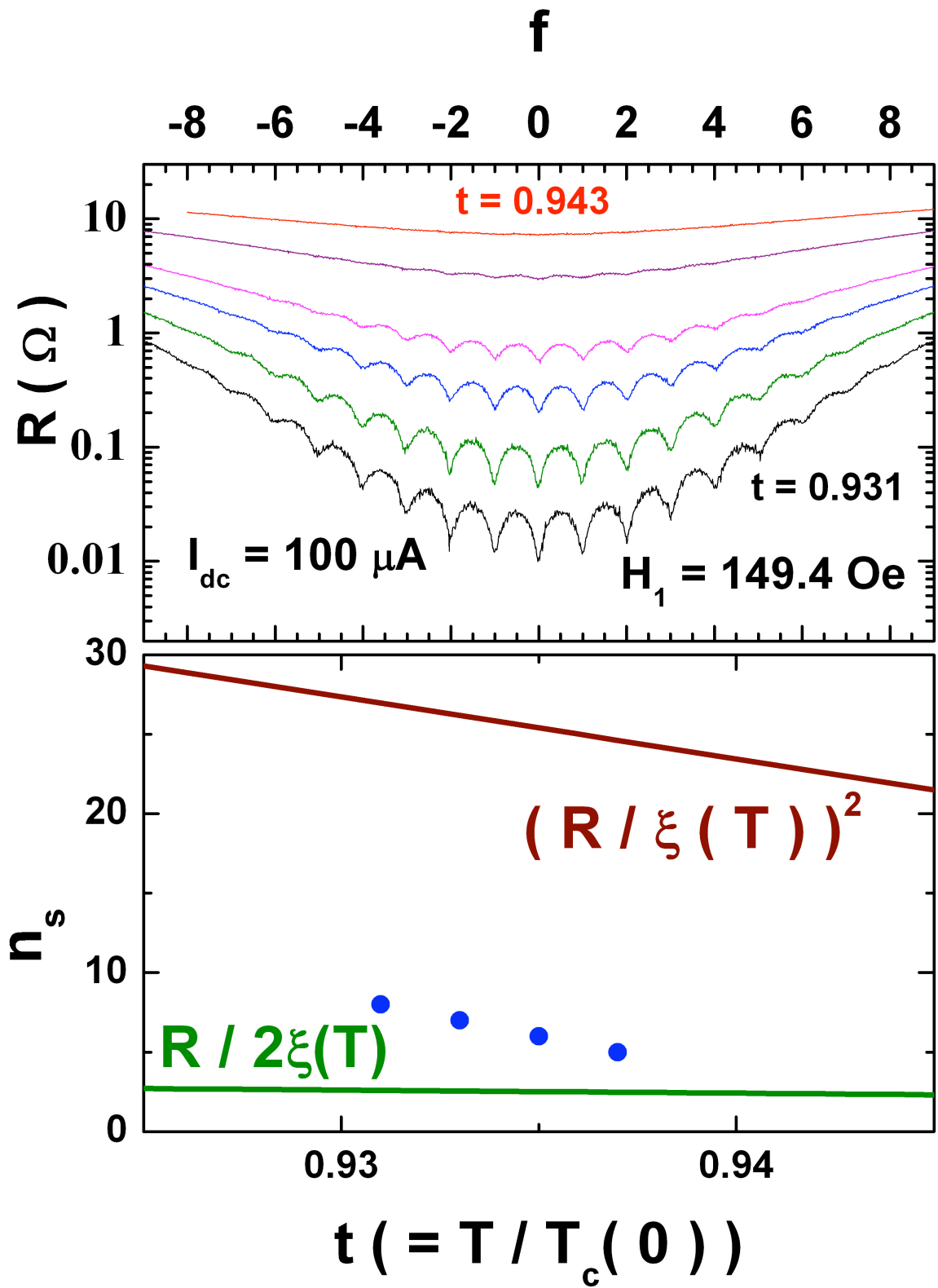

Fig. 3